# Residual Energy Based Cluster-head Selection in WSNs for IoT Application

Trupti Mayee Behera, Sushanta Kumar Mohapatra, Umesh Chandra Samal, Mohammad. S. Khan, Mahmoud Daneshmand, and Amir H. Gandomi

*Abstract*—**Wireless sensor networks (WSN) groups specialized transducers that provide sensing services to Internet of Things (IoT) devices with limited energy and storage resources. Since replacement or recharging of batteries in sensor nodes is almost impossible, power consumption becomes one of the crucial design issues in WSN. Clustering algorithm plays an important role in power conservation for the energy constrained network. Choosing a cluster head can appropriately balance the load in the network thereby reducing energy consumption and enhancing lifetime. The paper focuses on an efficient cluster head election scheme that rotates the cluster head position among the nodes with higher energy level as compared to other. The algorithm considers initial energy, residual energy and an optimum value of cluster heads to elect the next group of cluster heads for the network that suits for IoT applications such as environmental monitoring, smart cities, and systems. Simulation analysis shows the modified version performs better than the LEACH protocol by enhancing the throughput by 60%, lifetime by 66%, and residual energy by 64%.**

## I. INTRODUCTION

INTERNET of Things (IoT) is an ecosystem of interconnected devices and objects via the internet enabling them to send and receive data. It is an invisible but intelligent network that senses, controls and can be programmed [1], by employing embedded technology to communicate with each other. The IoT provides immediate access to information related to any device with high productivity and efficiency [2].

Till date, about 5 billion smart devices are already connected and by 2020 about 50 billion devices to be connected [1]. The number of people actually communicating may exceed the number of devices/machines connected to them virtually. This will generate huge traffic where humans may become the minority of generators and receivers of traffic [3]. This gives the reason for exploring IoT for various research areas owing to its challenges and opportunities

WSN acts as a medium that bridges the virtual digital world to the real world. Tiny sensors or actuators connected with each other are responsible for sensing and transferring the values to the Internet. WSN comprises sensor nodes deployed in a network field to monitor various physical and environmental parameters. The routing path of data from the sensing node to the sink node or base station (BS) should be designed in an energy efficient manner since recharging the sensor nodes is practically impossible [4].

Different from the ad-hoc network, WSN meant for IoT application faces numerous challenges in terms of a number of sensor nodes, hardware, mode of communication, battery power and computational cost to name a few. Apart from sensing, the sensors used in the IoT paradigm are assigned with additional functionalities and has to face new challenges in terms of QoS (quality of service), security and power management [5]. Some of these issues are addressed by adopting various technological changes in primitive protocols and schemes used for WSN. QoS requirements in IoT based WSN faces significant challenges like extreme resource content, redundancy in data, dynamic size of the network, less reliable medium, heterogeneous network, and multiple BS or sink nodes [6]. The key security issues in WSN includes data authenticity and confidentiality, data integrity and freshness in data [7].

Reduction of power consumption has always been a crux issue in designing WSNs. Recent research outcome has come up with different ideas to reduce energy and extend network longevity for proper utilization of resources. Routing algorithm plays a crucial role in the process. Clustering builds a hierarchy of clusters or groups of sensing nodes that collects and transfers the data to its respective cluster head (CH). The CH then groups the data and sends the fused to sink node or base station (BS) which acts as middleware between the end user and the network. Among the clustering algorithm, LEACH (Low Energy Adaptive Clustering Hierarchy) is a classical protocol that considers energy for hierarchical routing of data [8]. The network is grouped into clusters, and the sensor node transmits its data to the corresponding CH. The protocol randomly selects CHs in a stochastic manner for each round. The CH communicates with each node of the cluster called member nodes to collect the sensed data. The CH assigns TDMA (Time Division Multiple Access) schedules to its corresponding cluster member. The member node can transmit data during the allotted time slot. The data is then checked for redundancy and compressed before communicating with the sink node

The CHs directly communicate with BS in LEACH protocol; hence the power consumption in sending data from CH to BS will be more as compared to the communication between the CHs. As a result, the CHs will exhaust its energy within a short period of time. Multi-hop communication, on the other hand, can be helpful to overcome this problem, but still not effective in cases of small networks.

Electing a CH is a sophisticated job as various factors have to be considered for selection of best node in the cluster [9]. The factors include the distance between nodes, residual energy, mobility and throughput of each node.

LEACH algorithm enhances the lifetime of the network in comparison to direct or multi-hop transmission but still has many limitations. The election of cluster heads is done randomly which does not ensure proper distribution and optimal solution. The nodes with lesser energy have equal priority as that of those with a higher energy level to be elected as CH. So, when a node of low residual energy gets selected to serve as CH, it dies out quickly resulting in shorter network span [10].

The paper aims to select the CH considering important parameters like the initial energy, remaining energy of the individual node and the optimal number of CHs in the network. The modification is done in the classical LEACH algorithm. With the completion of each round, the residual energy of the non-CH nodes are checked, and the one with the higher energy level in comparison to others has a higher probability for CH selection for the current round. This would prevent the network to die out too early thereby enhancing the network lifetime.

The paper is structured as follows: Section II provides the required background and related work. Our proposed system model and algorithm are discussed in Section III. Section IV discusses the simulation results and analysis. Finally, the article is concluded in Section V.

## II. RELATED WORK

One of the major issues of IoT is to handle a large number of sensors that will be deployed, in terms of the cost of servicing and maintenance [11]. Further replacing sensor batteries which are already located in the network field can be a tedious job [12]. For instance, in case a sensor is to be deployed on a certain animal or species, it requires the battery of the sensor to outlive the animal which is far more achievable. This leads to another important challenge which is power management. Reliable end-to-end data transmission with proper congestion control and low packet loss ratio are some of the other major concerns in WSN [13].

The primary goal of any sensor network is to route the data assembled by sensors and forward it towards the BS. The simplest method to communicate data is direct transmission where the nodes have to direct its data to the base station or sink node. However, if the distance between sink and network is large, the node will die out quickly due to unnecessary energy consumption [14]. Clustering algorithm reduces the unwanted power consumption in delivering data to BS by grouping the network into clusters. Each cluster is assigned a CH that sends data to BS. An important phase in the clustering algorithm is the CH election process that should guarantee uniform energy distribution among the sensor nodes [15].

LEACH protocol has intensively been modified by researchers to improve the network performance. Technical researchers are contributing vigorously in enhancing existing algorithms for better performance of the IoT system [16]. An energy-efficient trust derivation method was discussed in [17] for WSN-based IoT networks. The scheme uses risk strategy analysis to reduce network overhead by deriving an optimal number of recommendations. The energy-aware scheme maintains adequate security and also reduces the latency of the network. A time-based CH selection is proposed in [18] called TB-LEACH that sets well-distributed clusters and enhances the lifetime by 20 to 30%. The distance between nodes and BS are considered for threshold based CH selection in [15] that improves lifetime by 10%.

Thein et al. in [14] have modified the probability for the selection of CH based on the residual energy of each node. The paper also considers the optimal value of CH but for fixed values like 1 and 6. The network lifetime enhances by 40-50%. Another CH selection method for aggregation of data is discussed in [19] that eliminates redundancy and enhances the network lifetime. The threshold value is modified by considering a hotness factor that defines the relative hotness of a particular sensor node to that of the network.

CH is selected using particle swarm optimization (PSO) in [20]. The criteria for selection have an objective function in terms of node degree, intra-cluster distance, residual energy, and a number of optimal CHs. The model performs better in terms of various network metrics in comparison to various routing protocols. PSO-ECHS is discussed in [21], where PSO based CH selection is made using parameters like node-to-node distance, distance to BS and residual energy. Another optimization technique called Grouped Grey Wolf Search Optimization is used in [22] for security-aware CH selection to improve the network lifetime.

An improvement of LEACH was proposed in [23] where residual energy plays an important role in CH election. A simple Multi-hop approach to LEACH was also studied, and it is found that both protocols perform better than LEACH by extending lifetime after a certain period of time. A non-probabilistic multi-criteria based CH selection was presented in [24] where the network is divided into separate zones. The CH or zone head is selected using the ANP (Analytical Network Process) decision tool. A set of parameters have been collected from where the best parameters have been selected for zone head selection.

The IoT, being a ubiquitous network, connects smart devices and objects to the cloud. WSN provides a platform for the collection and communication of data to monitor and control the physical world for the betterment of the society [25]. Communicating wireless technologies drains more power as compared to the devices meant to receive or sit idle. The rising number of smart devices connecting to the internet has made energy conservation a surplus parameter in IoT designing. Developing energy-efficient techniques for deployment of sensor networks have always been a challenging task for researchers. When incorporated with IoT, power becomes a more crucial issue owing to the number of devices being connected in large scale.

To maintain IoT standards, researchers have focussed on device energy-conserving techniques such as clustering where the choice of CH should be done judiciously. Various methods for efficient CH selection was studied from the above literature that enhances the network performance. However, important parameters like residual energy, initial energy and an optimum number of clusters in the network, have not been considered to the best of our knowledge for modification in the threshold value for CH selection.

## III. SYSTEM MODEL

The rapid increase in population density in urban areas requires modern infrastructures with suitable services to meet the requirements of the city inhabitants. Hence latest advances in communication technologies such as IoT has been in demand to provide a framework for the development of smart cities [26].

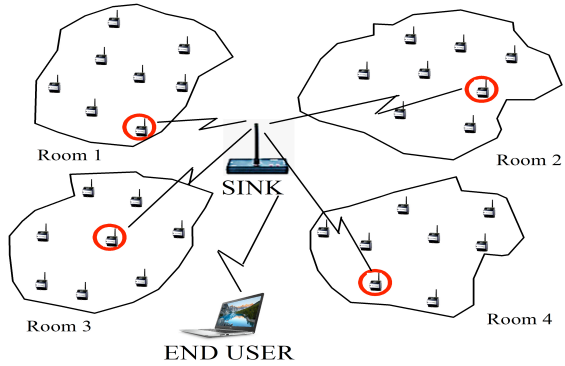

Fig. 1. Environment monitoring using IoT

This section presents an environmental monitoring scenario that uses WSN as an integral part of IoT. The nodes are grouped in four different rooms to form clusters as shown in Fig.1. Let there be eight sensor nodes in each room where only one node can become the CH (marked red) for each instant of time. The sink node collects data from the CHs of each room and sends the fused information to the end user.

For the system model, some reasonable assumptions have been adopted as follows:

- Nodes are static and homogeneous with initial energy 0.5J and are distributed in rooms to monitor variables such as humidity, temperature, sound, and luminosity.
- BS/sink is fixed and installed in the middle of the network

- Nodes are deployed randomly and transmit its data periodically
- Each room has a CH that communicates with the BS either in the single hop or multi-hop communication.
- The BS receives the data from each CH and spread it to the cloud.

The environment-monitoring applications [27] requires the proper routing of data so that the network energy can be used effectively. In case a node with lesser residual energy is elected as CH in one of the room, it will lead to the end of transmission of data from that room. As a result, the end user will not receive complete information for monitoring of environmental conditions. The communication model used in [28] shown in Fig. 2 has been considered to study the behavior of the proposed model.

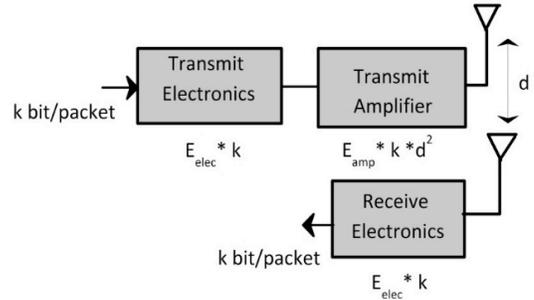

Fig. 2. Radio energy model

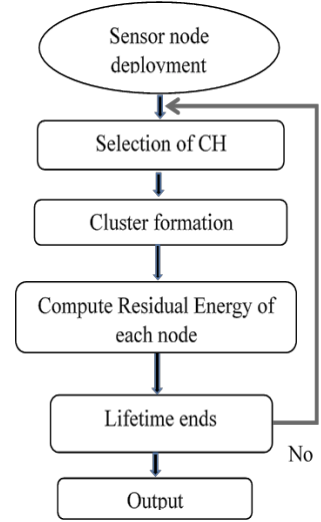

Fig. 3. Flowchart of LEACH protocol

For short distance communication as to that of between nodes and CHs within the room considers the free-space model and for longer distance transmission like between CHs and sink follows multi-path fading model [29]. For symmetrical propagation channel, the power consumption in transmitting '*k*' bits of data in a packet to a sensor placed '*d*' meters away can be written as [30]:

$$E_{Tx}(k,d) = E_{Tx\_ele}(k) + E_{Tx\_mp}(k,d) \qquad (1)$$

$$E_{Tx}(k,d) = \begin{cases} E_{ele} * k + E_{fs} * k * d^2, d \le d_0 \\ E_{ele} * k + E_{mp} * k * d^4, d > d_0 \end{cases} \quad (2)$$

$$E_{Rx}(k) = E_{Rx\_ele}(k) + kE_{ele} \quad (3)$$

Where $E_{ele}$ is the per bit energy consumption by receiver or transmitter. $E_{mp}$ is a transmission parameter for multi-path fading and $E_{fs}$ is for the free-space model [31].

LEACH is a primitive single-hop clustering protocol that saves an enormous amount of energy as compared to non-clustering algorithms [32]. Once the nodes are deployed, sensors group together to form clusters with one CH in each cluster for data aggregation. The protocol is implemented in rounds. Clusters are formed dynamically and the cluster heads are elected randomly. Each node in the cluster has equal probability to be elected as CH which intends to balance the energy dissipation. The residual energy is checked constantly by the sink until the lifetime ends, i.e. all nodes die out their battery power. The steps involved for each round in LEACH is depicted in the flowchart given in Figure 3.

With '$n$' sensor nodes in the field, let there be '$m$' clusters. In LEACH algorithm, $P$ represents the probability for a node to opt for CH. Before the start of the first round, each node '$i$' generates a random number between the interval [0,1]. If the number is found to be less than a threshold value $T(n)$ given by the equation (4), then that node will become CH for that round.

$$T(n) = \begin{cases} \dfrac{P}{1 - P(r \bmod \frac{1}{P})}; \text{ for all n } \acute{U}G \\ 0; \text{ Otherwise} \end{cases} \quad (4)$$

With each round, the CH changes based on the electing probability which indicates that all the nodes in the cluster have the same chance to be elected as CH irrespective of its residual energy. Equi-probable CH election process gives rise to the possibility of electing a CH with minimal residual energy which will die out quickly as compared to the one with comparative higher energy level. Therefore, the residual energy of each node is included in the equation of election probability of CH such that the nodes with higher energy level have a greater chance to be elected as CH. This in return ensures equal distribution of power in the network thus enhancing network lifetime.

In order to combat this problem, an advanced algorithm is proposed called R-LEACH. The algorithm is divided into rounds with each round consisting of cluster formation and steady-state stages.

## IV. PROPOSED WORK

The proposed protocol represents a hierarchical clustering algorithm that involves two stages: set-up and steady-state stages. In the initial set-up phase, the sensor nodes are deployed in the network and are sub-divided into clusters headed by a CH responsible for the collection of data from sensing nodes. The data is fused to reduce the volume by removing any redundant bits. Actual data routing occurs during the steady-state stage, where the collected data is forwarded to the BS by the CHs of the network.

**Set-up stage:**

For the first round, the clusters and CHs are formed using normal LEACH algorithm, where CHs are selected using equation (4). After data transfer, each node in the network expends some amount of energy which is different for every node. The expenditure of power depends on the distance separating the sending and receiving nodes represented as '$d$'. Hence for the next round, the CH is elected using a modified equation given as

$$T(n) = \begin{cases} \dfrac{P}{1 - P(r \bmod \frac{1}{P})} \times \dfrac{E_{residual}}{E_{initial}} \mathrm{k}_{opt}; \text{ for all n } \acute{U}G \\ 0; \text{ Otherwise} \end{cases} \quad (5)$$

Where $E_{residual}$ is the remaining energy level of the node and $E_{initial}$ is the initial assigned energy level. The optimal number of cluster $k_{opt}$ can be written as in [33].

$$k_{opt} = \sqrt{n/2\pi} \sqrt{\dfrac{E_{fs}}{E_{amp}d^4(2m-1)E_0 - mE_{DA}}} M \quad (6)$$

'$M$' represents the network diameter and $E_0$ is the initial energy supplied to each node.

Once the CHs for the current round are selected, they send their CH announcement information to member nodes in the respective clusters. The sensing nodes check the signal strength of the request message and decide the CHs it wants to join. The CH then broadcasts TDMA (Time Division Multiple Access) schedules for the member nodes to transmit data in different time slots to avoid data collision. The process then continues for the rest of the rounds till all the nodes in the network exhaust all its energy.

**Steady-state stage:**

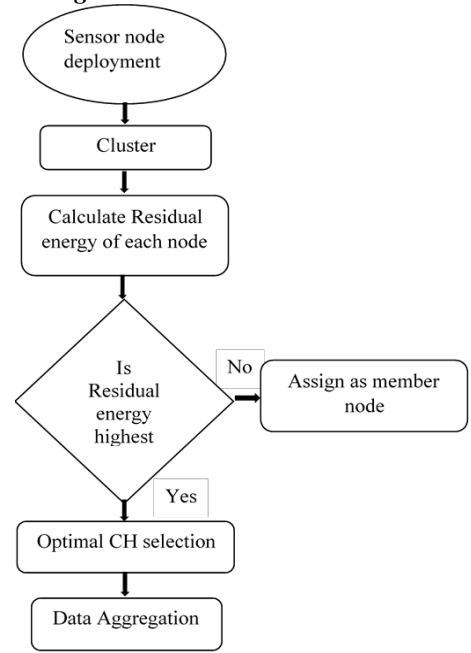

Fig. 4.  Flowchart of R-LEACH

During the time slot assigned to each node, transmission of data to CHs occurs. Only the transmitting node remains active and all other nodes in the cluster will turn-off its radio to save energy. After all the nodes in the cluster have finished transferring data, the CH will start processing the data. The CH receives and then aggregates the data to remove any redundancy and compress the information as much as possible for fair utilization of bandwidth. The CHs then forwards the data to the sink or BS in either single-hop or multi-hop communication. The entire process is depicted in a flowchart as shown in Fig. 4.

## V. Simulations

The network parameters considered for MATLAB simulation for the network model are described in Table 1. The packet size is considered to be 4000 bits. 100 nodes are deployed randomly with BS placed in the center of the network area as shown in Fig. 5.

TABLE I
SIMULATION PARAMETERS

| Parameters | Value |
|---|---|
| Network diameter | 100 meters$^2$ |
| Total number of nodes (n) | 100 nodes |
| Total network energy ($E_0$) | 0.5 J |
| Energy dissipation: receiving ($E_{amp}$) | 0.0013 pJ/bit/m$^4$ |
| Energy dissipation: free space model ($E_{fs}$) | 10 pJ/bit/m$^2$ |
| Energy dissipation: power amplifier ($E_{amp}$) | 100 pJ/bit/m$^2$ |
| Energy dissipation: aggregation ($E_{DA}$) | 5 nJ/bit |

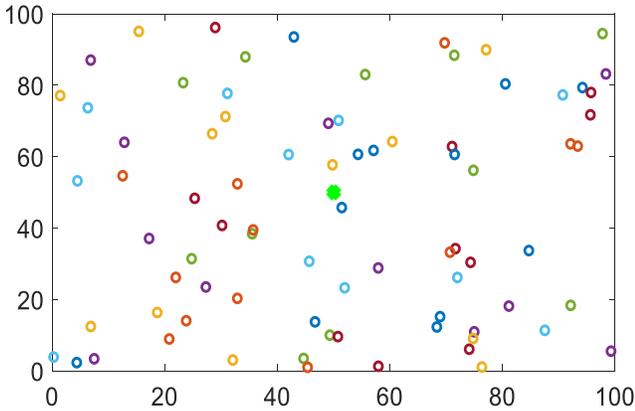

Fig. 5. Node Deployment

### A. Network Analysis

The simulation result in Fig.6 shows the network life for both the LEACH and R-LEACH protocols.

The first node for LEACH dies out at 1092 round whereas for R-LEACH it is at 1382 rounds. Similarly, the last node for LEACH dies out at 1510 rounds whereas for R-LEACH it is at 2474 rounds. LEACH protocol assumes CHs dissipates the same energy for each round that leads to inefficient CH selection and affects the network lifespan. R-LEACH selects CHs considering the residual energy of nodes and an optimal number of clusters together thereby enhancing the network lifetime to more number of rounds.

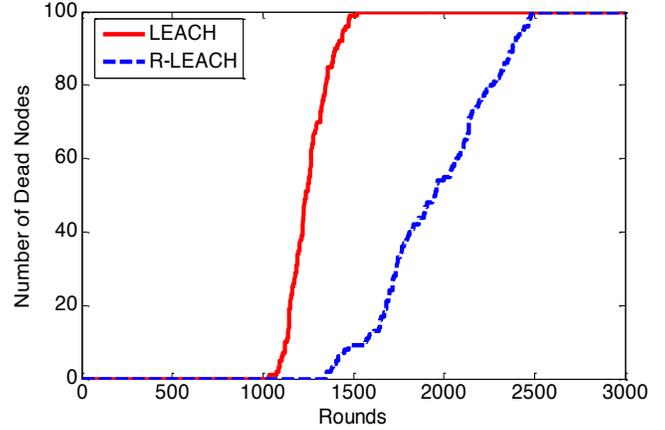

Fig. 6. Network Lifetime

The number of actual data packets sent to sink is shown in Fig.7. Since the CHs are selected based on the remaining energy of each node, it effectively reduces the energy dissipation in transferring data. As a result, data transmission frequency increases and more packets are successfully transmitted to the BS as compared to that in LEACH protocol. The average energy expenditure of the network is shown in Fig. 8. The residual energy depletes faster in LEACH than that of R-LEACH. Since the energy for modified LEACH depletes at a slow rate, the network lifetime also extends to more number of rounds.

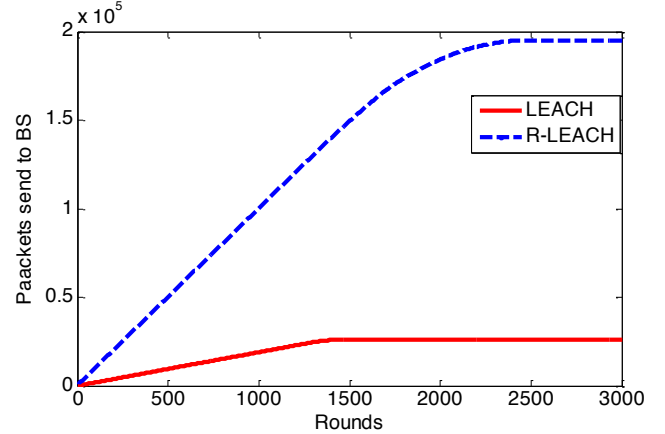

Fig. 7. Packets to BS

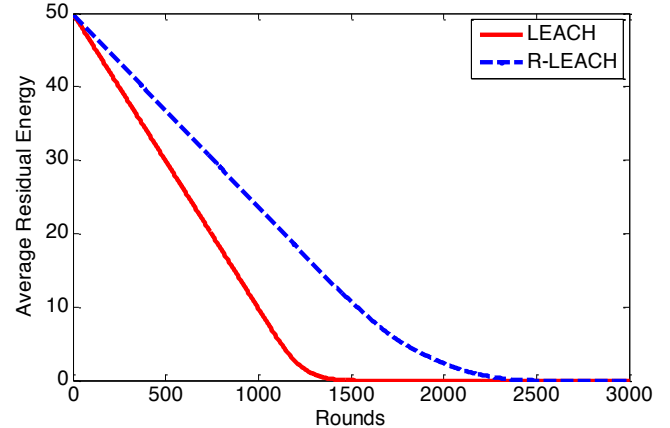

Fig. 8. Average Residual energy

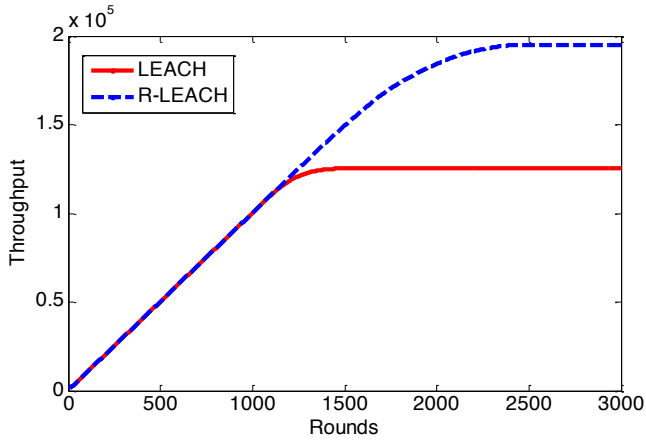

Fig. 9. Throughput

Throughput represents the ratio between actually transmitted data packets to the successfully received data at BS or sink. The higher the ratio, the better is the performance. Fig. 9 show the graph of the throughput of the two protocols. It is clear that due to the modification in the threshold value of CH selection, the throughput is increased by 60% for R-LEACH. Hence, it can be concluded that the modified protocol performs better than LEACH protocol and can be used extensively for homogeneous networks.

### B. Variation of energy and its effect on Network metrics

Network Stability and Network Lifetime are two important parameters to deliver good network performance in an IoT-based environment consistently. Network Stability is the time from the beginning of the network till the death of the first node (FND) and Network Lifetime is determined by the time elapsed between FND and LND (death of the last node) in the network. To study the behavior of proposed R-LEACH protocol with respect to these metrics, we analyze the parameters like FND for network stability, LND for network lifetime and HND in the network. For a fair comparison with recent methods proposed for IoT based WSN like IGHND [24], GHND [34], CBDAS [35], we consider the packet size to be 2000. The initial energy $E_0$ of 0.25, 0.5 and 1 J are considered for network analysis.

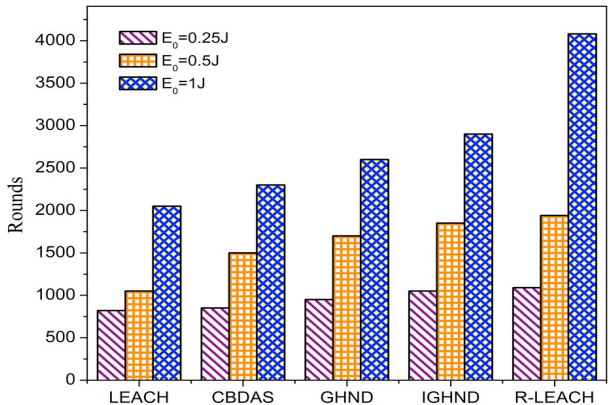

Fig. 10. Network Stability

The result in Fig.10 shows R-LEACH provides a more stable network for all the energy values than other routing protocols for an IoT based sensor network. LEACH does not consider energy level for the threshold value of CH selection; hence the first node dies out after less number of rounds in all cases. CBDAS and GHND are grid-based WSNs that consider few parameters for CH selection and perform better than LEACH protocol. IGHND divides the network into zones and selects a zone head considering parameters like residual energy, average distance, and priority of a node. None of the methods considers the optimum value of clusters in the network resulting in inefficient CH selection. R-LEACH takes into account both the important parameters which result in an improved stable network.

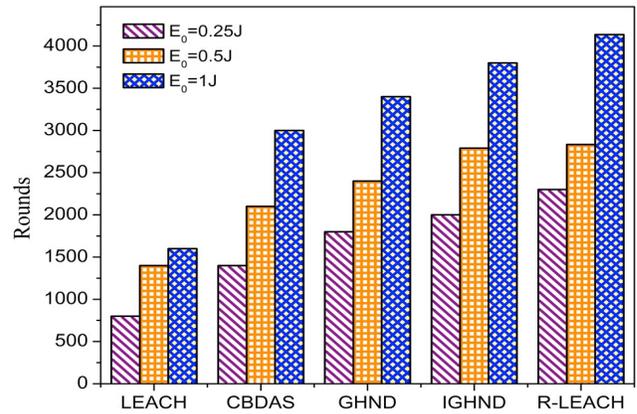

Fig. 11. Half Node Dead

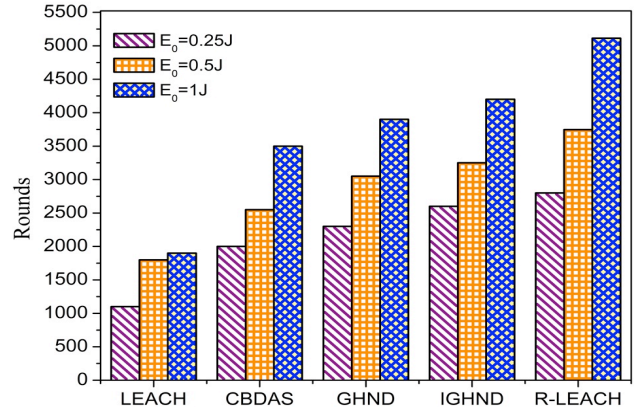

Fig.12. Network Lifetime

The proposed R-LEACH model maximizes the network lifespan by covering more rounds of operation and transmitting more packets to the BS. R-LEACH outperforms LEACH, CBDAS, GHND, and IGHND in terms of HND and FND for all energy values because it selects stable nodes as CH. In LEACH, the CH selection is done in a randomized manner leading to a short span of the network. CBDAS consumes extra energy in chain formation and data transmission from header to rest of the nodes adding extra load to the battery life. GHND and IGHND consider multiple parameters for zone head selection but does not consider the number of zones or cluster in the network which effects the network lifetime.

## VI. Conclusion

Since energy and lifetime are two major constraints in designing any routing protocol for WSN, much research has been done to achieve the goal. Choosing an energy-efficient routing algorithm that distributes the load in the network evenly is a challenging process. LEACH protocol ensures an adaptive algorithm but still has some limitations. A modified CH selection algorithm has been suggested in this paper that aims to extend the network lifetime by controlling the energy dissipation in the network. The enhanced routing process can be used effectively in scenarios like environmental monitoring using IoT as the protocol delivers a better result for homogeneous networks in comparison to LEACH. Simulation result shows improved network performance for metrics such as residual energy, packets sent to BS, throughput and lifetime. The current work can be extended by considering more parameters for CH selection in a network with mobile nodes that changes its position frequently. The proposed model can also be tested on different realistic scenarios for a WSN based IoT system.

## References

[1] J. Chase, "The evolution of the internet of things," *Texas Instruments*, 2013.

[2] D. Bandyopadhyay and J. Sen, "Internet of things: Applications and challenges in technology and standardization," *Wirel. Pers. Commun.*, vol. 58, no. 1, pp. 49–69, 2011.

[3] L. Tan and N. Wang, "Future internet: The internet of things," in *2010 3rd International Conference on Advanced Computer Theory and Engineering (ICACTE)*, 2010, vol. 5, pp. V5--376.

[4] W. B. Heinzelman, A. P. Chandrakasan, and H. Balakrishnan, "An application-specific protocol architecture for wireless microsensor networks," *IEEE Trans. Wirel. Commun.*, vol. 1, no. 4, pp. 660–670, 2002.

[5] J. Gubbi, R. Buyya, S. Marusic, and M. Palaniswami, "Internet of Things (IoT): A vision, architectural elements, and future directions," *Futur. Gener. Comput. Syst.*, vol. 29, no. 7, pp. 1645–1660, 2013.

[6] B. Bhuyan, H. K. D. Sarma, N. Sarma, A. Kar, and R. Mall, "Quality of service (QoS) provisions in wireless sensor networks and related challenges," *Wirel. Sens. Netw.*, vol. 2, no. 11, p. 861, 2010.

[7] Q. Jing, A. V Vasilakos, J. Wan, J. Lu, and D. Qiu, "Security of the Internet of Things: perspectives and challenges," *Wirel. Networks*, vol. 20, no. 8, pp. 2481–2501, 2014.

[8] W. R. Heinzelman, A. Chandrakasan, and H. Balakrishnan, "Energy-efficient communication protocol for wireless microsensor networks," in *Proceedings of the 33rd Annual Hawaii International Conference on System Sciences*, 2000, p. 10 pp. vol.2-.

[9] M. Chatterjee, S. K. Das, and D. Turgut, "An on-demand weighted clustering algorithm (WCA) for ad hoc networks," in *Global Telecommunications Conference, 2000. GLOBECOM'00. IEEE*, 2000, vol. 3, pp. 1697–1701.

[10] J. Xu, N. Jin, X. Lou, T. Peng, Q. Zhou, and Y. Chen, "Improvement of LEACH protocol for WSN," in *Fuzzy Systems and Knowledge Discovery (FSKD), 2012 9th International Conference on*, 2012, pp. 2174–2177.

[11] Y.-K. Chen, "Challenges and opportunities of the internet of things," in *17th Asia and South Pacific Design Automation Conference*, 2012, pp. 383–388.

[12] A. Ali, Y. Ming, T. Si, S. Iram, and S. Chakraborty, "Enhancement of RWSN Lifetime via Firework Clustering Algorithm Validated by ANN," *Information*, vol. 9, no. 3, p. 60, 2018.

[13] C. Wang, K. Sohraby, B. Li, M. Daneshmand, and Y. Hu, "A survey of transport protocols for wireless sensor networks," *IEEE Netw.*, vol. 20, no. 3, pp. 34–40, 2006.

[14] M. C. M. Thein and T. Thein, "An energy efficient cluster-head selection for wireless sensor networks," in *Intelligent systems, modelling and simulation (ISMS), 2010 international conference on*, 2010, pp. 287–291.

[15] S. H. Kang and T. Nguyen, "Distance based thresholds for cluster head selection in wireless sensor networks," *IEEE Commun. Lett.*, vol. 16, no. 9, pp. 1396–1399, 2012.

[16] J. A. Stankovic, "Research directions for the internet of things," *IEEE Internet Things J.*, vol. 1, no. 1, pp. 3–9, 2014.

[17] J. Duan, D. Gao, D. Yang, C. H. Foh, and H.-H. Chen, "An energy-aware trust derivation scheme with game theoretic approach in wireless sensor networks for IoT applications," *IEEE Internet Things J.*, vol. 1, no. 1, pp. 58–69, 2014.

[18] H. Junping, J. Yuhui, and D. Liang, "A time-based cluster-head selection algorithm for LEACH," in *Computers and Communications, 2008. ISCC 2008. IEEE Symposium on*, 2008, pp. 1172–1176.

[19] K. Maraiya, K. Kant, and N. Gupta, "Efficient cluster head selection scheme for data aggregation in wireless sensor network," *Int. J. Comput. Appl.*, vol. 23, no. 9, pp. 10–18, 2011.

[20] B. Singh and D. K. Lobiyal, "A novel energy-aware cluster head selection based on particle swarm optimization for wireless sensor networks," *Human-Centric Comput. Inf. Sci.*, vol. 2, no. 1, p. 13, 2012.

[21] P. C. S. Rao, P. K. Jana, and H. Banka, "A particle swarm optimization based energy efficient cluster head selection algorithm for wireless sensor networks," *Wirel. networks*, vol. 23, no. 7, pp. 2005–2020, 2017.

[22] A. Shankar, N. Jaisankar, M. S. Khan, R. Patan, and B. Balamurugan, "Hybrid model for security-aware cluster head selection in wireless sensor networks," *IET Wirel. Sens. Syst.*, 2018.

[23] F. Xiangning and S. Yulin, "Improvement on LEACH protocol of wireless sensor network," in *Sensor Technologies and Applications, 2007. SensorComm 2007. International Conference on*, 2007, pp. 260–264.

[24] H. Farman *et al.*, "Multi-criteria based zone head selection in Internet of Things based wireless sensor networks," *Futur. Gener. Comput. Syst.*, 2018.

[25] S. Kallam, R. B. Madda, C.-Y. Chen, R. Patan, and D. Cheelu, "Low energy-aware communication process in IoT using the green computing approach," *IET Networks*, vol. 7, no. 4, pp. 258–264, 2017.

[26] J. Jin, J. Gubbi, S. Marusic, and M. Palaniswami, "An information framework for creating a smart city through internet of things," *IEEE Internet Things J.*, vol. 1, no. 2, pp. 112–121, 2014.

[27] E. Souto, G. Guimarães, G. Vasconcelos, M. Vieira, N. Rosa, and C. Ferraz, "A message-oriented middleware for sensor networks," in *Proceedings of the 2nd workshop on Middleware for pervasive and ad-hoc computing*, 2004, pp. 127–134.

[28] J. Li and P. Mohapatra, "An analytical model for the energy hole problem in many-to-one sensor networks," in *IEEE vehicular technology conference*, 2005, vol. 61, no. 4, p. 2721.

[29] B. Elbhiri, R. Saadane, D. Aboutajdine, and others, "Developed Distributed Energy-Efficient Clustering (DDEEC) for heterogeneous wireless sensor networks," in *I/V Communications and Mobile Network (ISIVC), 2010 5th International Symposium on*, 2010, pp. 1–4.

[30] D. S. Kim and Y. J. Chung, "Self-Organization Routing Protocol Supporting Mobile Nodes for Wireless Sensor Network," in *Computer and Computational Sciences, 2006. IMSCCS '06. First International Multi-symposiums on*, 2006, vol. 2, pp. 622–626.

[31] T. M. Behera, U. C. Samal, and S. K. Mohapatra, "Energy Efficient Modified LEACH Protocol for IoT Application," *IET Wirel. Sens. Syst.*, vol. 8, no. 5, pp. 223 – 228, 2018.

[32] Y. Li, N. Yu, W. Zhang, W. Zhao, X. You, and M. Daneshmand, "Enhancing the performance of LEACH protocol in wireless sensor networks," in *Computer Communications Workshops (INFOCOM WKSHPS), 2011 IEEE Conference on*, 2011, pp. 223–228.

[33] S. Hussain and A. W. Matin, "Energy efficient hierarchical cluster-based routing for wireless sensor networks," *Jodrey Sch. Comput. Sci. Acadia Univ. Wolfville, Nov. Scotia, Canada, Tech. Rep.*, pp. 1–33, 2005.

[34] H. Farman, H. Javed, J. Ahmad, B. Jan, and M. Zeeshan, "Grid-based hybrid network deployment approach for energy efficient wireless sensor networks," *J. Sensors*, vol. 2016, 2016.

[35] Y.-K. Chiang, N.-C. Wang, and C.-H. Hsieh, "A cycle-based data aggregation scheme for grid-based wireless sensor networks," *Sensors*, vol. 14, no. 5, pp. 8447–8464, 2014.

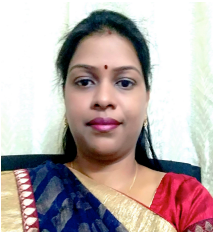

**T. M. Behera** got her B.Tech degree in Electronics & Comm. Engineering from Biju Pattnaik University of Technology in 2007. She received her M.Tech in Communication System from KIIT University in 2012. She has 10 years of teaching experience and is currently an Assistant Professor in SOEE at KIIT University, Bhubaneswar. Her research area broadly includes Wireless Sensor Networks and its application in IoT.

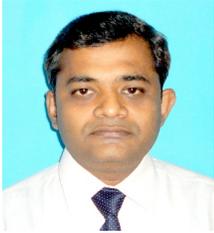

**Dr. S. K. Mohapatra** received his Ph.D. from National Institute of Technology, Rourkela, in the year 2016. He is currently an Assistant Professor in SOEE, KIIT University, Bhubaneswar and having 18 years of teaching experience. His research interests include Modeling and Simulation of Nanoscale Devices and its application in IoT. Energy-efficient Wireless Sensor Networking, Adhoc Networks, Metamaterial absorbers in THz application, UWB-MIMO and Reconfigurable Antenna. He has been a part of committee member of various international conferences, Editorial Board Member and Reviewer of international journals. He is a life member of ISTE, IETE, CSI, OITS, and member of IEEE.

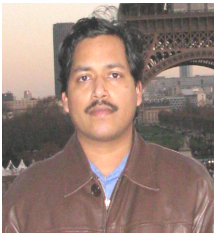

**Dr. U. C. Samal**, received his B.E. degree in Electrical Engineering from VSSUT, Burla, Sambalpur, India in 2003 and M. Tech. degree in Electronic System and Comm. Engineering from National Institute of Technology, Rourkela, in 2006. He obtained his Ph. D. degree from the Department of Electrical Engineering, Indian Institute of Technology (IIT), Kanpur, India in 2015. His area of specialization lies in wireless communication systems, signal processing techniques for communication, 5G wireless communication technologies cognitive radio and wireless sensor networks. He has 3 years of industry and 4 years of teaching experience. Currently, he is working as Assistant Professor (II) at KIIT University, Bhubaneswar, Odisha.

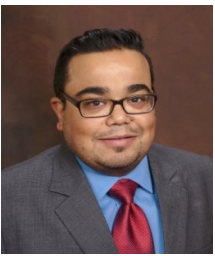

**Dr. Mohammad S. Khan** is currently an Assistant Professor of Computing at East Tennessee State University and the director of Network Science and Analysis Lab (NSAL). He received his M.Sc. and Ph.D. in Computer Science and Computer Engineering from the University of Louisville, Kentucky, USA, in 2011 and 2013, respectively. His primary area of research is in ad-hoc networks, network tomography, connected vehicles, and vehicular social networks. He currently serves as Co-Editor-in-Chief of *International Journal of Grid and High-Performance Computing* (IJGHPC) and as an associate editor of IEEE *Access*. He has been on technical program committees of various international conferences and technical reviewer of various international journals in his field. He is a member of IEEE.

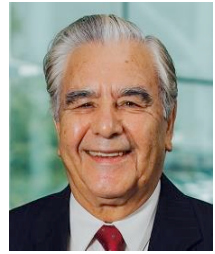

**Dr. Mahmoud Daneshmand** is Co-Founder and Professor of Department of Business Intelligence & Analytics; and Professor of Department of Computer Science at Stevens Institute of Technology. He has more than 40 years of Industry & University experience as Professor, Researcher, Assistant Chief Scientist, Executive Director, Distinguished Member of Technical Staff, Technology Leader, Chairman of Department, and Dean of School at: Bell Laboratories; AT&T Shannon Labs–Research; University of California, Berkeley; University of Texas, Austin; Sharif University of Technology; University of Tehran; New York University; and Stevens Institute of Technology. He received his Ph.D. and M.S. degrees in Statistics from the University of California, Berkeley; M.S. and B.S. degrees in Mathematics from the University of Tehran.

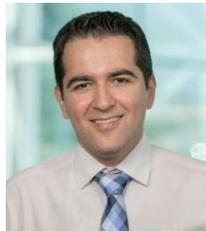

**Dr. Amir H. Gandomi** is an Assistant Professor of Analytics & Information Systems at School of Business, Stevens Institute of Technology. Prior to joining Stevens, Dr. Gandomi was a distinguished research fellow in headquarter of BEACON NSF center located at Michigan State University. He received his Ph.D. in engineering and used to be a lecturer in several universities. Dr. Gandomi has published over one hundred and thirty journal papers and four books. Some of those publications are now among the hottest papers in the field and collectively have been cited more than 11,000 times (h-index = 52). He has been named as Highly Cited Researcher (top 1%) for two consecutive years, 2017 and 2018, and one of the world's most influential scientific minds. Dr. Gandomi is currently ranked 20th in GP bibliography among more than 11,000 researchers. He has also served as associate editor, editor and guest editor in several prestigious journals and has delivered several keynote/invited talks. Dr. Gandomi is part of a NASA technology cluster on Big Data, Artificial Intelligence, and Machine Learning. His research interests are global optimization and (big) data mining using machine learning and evolutionary computations in particular.